# Upper critical fields of $MgB_2$ thin films

V.Ferrando[a], P.Manfrinetti[b], D.Marré[a], M.Putti[a], C.Tarantini[a], C.Bernini[a] and C.Ferdeghini[a] [*]

[a] *INFM-LAMIA and DIFI, Università di Genova, via Dodecaneso 33, 16146 Genova Italy*

[b] *INFM-LAMIA and DICCI, via Dodecaneso 31, 16146 Genova Italy*

**Abstract**

Critical fields of four $MgB_2$ thin films with a normal state resistivity ranging from 5 to 50 μΩcm and $T_c$ from 29.5 to 38.8 K were measured up to 28 T. $H_{c2}(T)$ curves present a linear behavior towards low temperatures. Very high critical field values have been found, up to 24 T along the c-axis and 57 T in the basal plane not depending on the normal state resistivity values. In this paper, critical fields will be analyzed taking into account the multiband nature of $MgB_2$; we will show that resistivity and upper critical fields can be ascribed to different scattering mechanisms.

*Keyword*s: $MgB_2$; thin films; multiple gap superconductivity

[*] Corresponding author: Tel.: +39-010-3536282; fax: +39-010-311066; e-mail: ferdeghini@fisica.unige.it.

Since the discovery of its superconductivity [1], Magnesium Diboride showed a large number of intriguing properties. It is now well clarified [2] that it is a two-band system with two distinct energy gaps; the larger associated to the two-dimensional $\sigma$ bands and the smaller to the near isotropic three-dimensional $\pi$ bands. In this scenario, most of the superconducting properties of $MgB_2$ can be explained [3-5], but the influence of these two bands on the upper critical fields and their anisotropy $\gamma$ is not clear yet. Furthermore, a strong difference between single crystals and thin films has been remarked, the formers showing low resistivity, low critical fields and $\gamma \sim 6$ [6,7] and the latter having a big spread in $T_c$ and resistivity values, high critical fields and lower anisotropy [8-12]. In general the higher $H_{c2}$ values in thin films were related to the higher resistivity in the normal state induced by disorder. In this paper, we will show, instead, how thin film with $\rho(40\ K)$ comparable with those of single crystals can show very high $H_{c2}$ values. Critical fields and resistivity in fact can be determined by different scattering mechanisms, thus explaining why thin films, in which $\rho(40\ K)$ varies of one order of magnitude can present very similar $H_{c2}$ values. In order to discuss this topic, we selected four different thin films grown by Pulsed Laser Ablation, whose properties are summarized in table I

Details about the deposition technique are reported elsewhere [10]. Due to their different purity, a complete analysis of this set of samples can be very helpful for our scope. In thin films indeed, the disorder on one hand enhances the resistivity in the normal state and on the other, increasing the inter-band scattering, reduces $T_c$ [13]. In any case the inter-band scattering remains negligible with respect to the intra-band ones [14]. All the four samples analyzed show, in $\theta$-$2\theta$ patterns, intense (00l) peaks indicating a strong *c*-axis orientation of the phase. Only in film 4, (101) reflection, which is the most intense in powders, seems to be detectable, even if with very low intensity; this indicates a not perfect orientation of the film. The films show $T_c$ ranging from 29.5 K to 38.8 K and $\rho(40\ K)$ in the range 5-50 $\mu\Omega$cm. Electrical resistivity measurements were performed in magnetic field up to 28 T at Grenoble High Magnetic Field Laboratory. $H_{c2}(T)$ curves parallel and perpendicular to the *ab* planes ($H_{c2}^{//}$ and $H_{c2}^{\perp}$) have been obtained using the criterion of 90% of normal state resistivity. They are reported in figure 1.

Despite the very different characteristics of the four films, a general trend in $H_{c2}^{\perp}(T)$ curves can be observed a part from film 4, whose $H_{c2}^{\perp}$ seems to be considerably higher than the others, probably due to a not perfect *c*-orientation, which influences only upper critical field along the *c*-axis, the parallel one being higher. Some interesting features can be noted in these data. First, the $H_{c2}(T)$ curves show a linear behavior towards low temperatures even at the lowest temperatures we measured, so that any saturation is observed. This is well evident in particular in perpendicular orientation, where the magnetic field we can apply is strong enough to determine $H_{c2}$ down to 2 K, allowing a reasonable estimation of $H_{c2}^{\perp}(0)$ by linear extrapolation. BCS extrapolation at 0 K strongly underestimates the real $H_{c2}(0)$. For film 1 for example, the calculated BCS extrapolation value of $H_{c2}^{\|}(0)=22T$ and $H_{c2}^{\perp}(0)=8.75$ Tesla are already reached at 13 and 10 K respectively. Moreover, the critical fields values are similar even if normal state resistivity values vary of one order of magnitude. This is different from what predicted from BCS theory (and well followed by low temperature superconductors) i.e. that the slope $dH_{c2}/dT$ at $T_c$ is proportional to residual resistivity so an increase in $\rho$ should proportionally increase $H_{c2}(0)$. In the whole set of our data, instead, we cannot observe a clear $H_{c2}$ dependence on $\rho$. From $H_{c2}$ data of figure 1, it is possible to estimate the anisotropy factors $\gamma=H_{c2}^{\|}/H_{c2}^{\perp}$ for all the films. Their temperature dependences are reported in the inset of figure 1. All $\gamma(T)$ curves have the same behavior, $\gamma$ decreasing when the temperature increases and all the values are in the range between 3 and 3.5 at the lowest temperatures, except the case of film 4 where $\gamma$ is 2.3, probably due to the not perfect *c*-orientation. Considering that the $\gamma$ curves seem to saturate at low temperature, it is reasonable to use the $\gamma$ values at the lowest temperature to estimate $H_{c2}^{\|}$ from $H_{c2}^{\perp}$ values (see Table I). The so calculated $H_{c2}^{\|}$ values fall in the narrow range between 42 and 57 T despite the differences in normal state resistivity values. To explain this phenomenology, the multiband nature of $MgB_2$ must be taken into account, as in [15]. Following this model, it is possible to calculate $H_{c2}(0)$ from the most effective scattering mechanism. In all our films, diffusivity in the π band, $D_\pi$, resulted to be higher than the one in σ band, $D_\sigma$, indicating that the scattering in the σ band determines upper critical fields. An accurate analysis of resistivity data of this set of samples [14] on the contrary, showed that resistivity is dominated by the scattering in the π band. This fact can explain why, differently from what predicted from BCS theory, films with resistivity similar to that of single crystals can present upper critical field values considerably high. This work was supported by the European Community through "Access to Research Infrastructure action of the Improving Human Potential Programme".

|  | *FILM 1* | *FILM 2* | *FILM 3* | *FILM 4* |
|---|---|---|---|---|
| substrate | $Al_2O_3$ c-cut | MgO (111) | MgO (111) | $Al_2O_3$ c-cut |
| $T_c$, K | 29.5 | 32 | 33.9 K | 38.8 K |
| $\Delta T_c$, K | 2 | 1.5 K | 1.1 K | 1 K |
| RRR | 1.2 | 1.3 | 1.5 | 2.5 |
| $\rho$(40K), $\mu\Omega$cm | 40 | 50 | 20 | 5 |
| $\gamma$ | 3 | 3.5 | 3 | 2.3 |
| $H_{c2}(0) \perp ab$, T | 14.2 | 15.5 | 16.8 | 24.6 |
| $H_{c2}(0) //ab$, T | 42 | 54 | 50 | 57 |

Table I: Main properties of the four films. $T_c$ is the onset of the transition (90% of $\rho$(40 K)) and $\Delta T_c$ is calculated between the 90 % and 10 % of $\rho$(40 K). Accuracy on the resistivity is of 20% due to the uncertainty in thickness determination..

**Figure Caption**

Fig.1: Critical fields parallel and perpendicular to the basal plane for the four samples as a function of the reduced temperature. In the inset, anisotropy factors $\gamma=H_{c2}(\theta=0°)/H_{c2}(\theta=90°)$ as a function of reduced temperature.

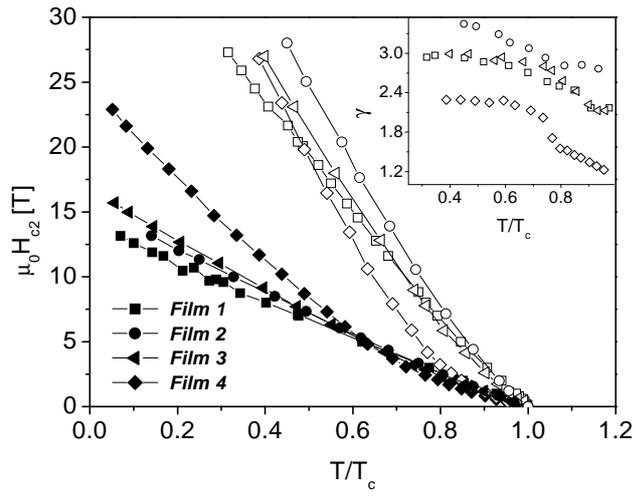

Fig.1